# A Generalized-Impedance Based Stability Criterion for Three-Phase Grid-Connected Voltage Source Converters

Huanhai Xin, Ziheng Li, Wei Dong, Zhen Wang, and Leiqi Zhang

*Abstract*—The output impedance matrices of three-phase grid-connected voltage source converters (VSCs) are widely used in power system stability analysis. Regardless of how the impedance is modeled, there always exist coupling terms in the impedance matrix, which makes the system a multi-input-multi-output (MIMO) system. As a practical approximation, the coupling terms are generally omitted and the impedance-based stability criterion for the resultant single-input-single-output (SISO) system can be applicable. However, such handling may result in analytical errors. This letter proposes a new stability criterion based on an equivalent SISO system by introducing the concept of generalized- impedances, which can completely keep the coupling terms. Further, the effects of the phase-locked-loop (PLL) parameters on system stability are studied based on the proposed criterion. The effectiveness of the proposed criterion is verified by a hardware-in-the-loop (HIL) simulation based on a RT-LAB platform.

## I. INTRODUCTION

GRID-CONNECTED voltage source converters (VSCs) are widely used for renewable energy integration and high-voltage DC transmission systems [1]. However, the interaction between the power electronic controllers and the transmission lines can lead to power oscillations, which poses new challenges for power system stability [2-5].

When analyzing the oscillation problems introduced by power electronic devices, the impedance-based methods are widely used, in which a grid-connected VSC is modeled as an output impedance connected to a voltage/current source and the grid is represented similarly [2-7]. In a three-phase system, the impedance of the VSC or the grid can be mathematically written as a matrix, which relates the voltage vector to the current vector. Currently, most impedance-based methods for grid-tied VSCs can be classified into two categories according to the reference frames where the impedance matrices are formulated: the synchronous frame (dq frame) based methods [2-3] and the stationary frame (abc frame) based methods [4-5].

In the dq frame, there usually exist coupling terms (non-diagonal elements) in the impedance matrices of the VSCs and the grid, which makes the system a MIMO system. Usually, the generalized Nyquist stability criterion (GNSC) is used for stability performance analysis in these methods [2-3]. Similarly, in the abc frame, the coupling terms also exist between the positive and the negative sequence impedances in the VSC system and the grid. Although the two sequence impedances of a symmetrical grid can be strictly decoupled (the impedance matrix is diagonal), the impedance matrix of the VSC usually has coupling terms [5]. As a practical approximation, the coupling terms are generally omitted and the impedance-based stability criterion for the resultant SISO system can be applicable [4, 6]. Unfortunately, ignoring the coupling terms may lead to inaccurate analysis results in a mirror frequency coupled system [5, 7].

This letter presents a generalized-impedance based stability criterion (GISC) for three-phase grid-connected VSCs. By mathematically manipulating the characteristic equation, the system can be transformed into an equivalent SISO system, which can be regarded as a generalized-impedance based series circuit. In essence, the small-signal instability of the system can be explained as a resonance of the equivalent circuit and evaluated by the classical Nyquist stability criterion.

## II. GRID-CONNECTED VSC MODEL

The grid-connected VSC considered in this letter is shown in Fig. 1. The converter uses a conventional LCL filter. The grid-side network includes $C_f$ and $L_{Line}$, where $C_f$ is the filter capacitor, $L_{Line}$ denotes the total inductance of the filter and the transmission line. The converter is controlled by a dual-loop vector controller based on a PLL. The current controller of the VSC has voltage feed-forward (VFF) and decoupling terms. The positions of the phasors and the reference frames of the system are shown in Fig. 2, where the dq frame represents the converter-side rotating frame introduced by the PLL and the xy frame represents the grid-side synchronous rotating frame [2]. In steady state, the dq frame is aligned with the xy frame.

This paper focuses on the oscillations induced by the PLL and the inner current loop (similar to [5]). The following assumptions are considered to simplify the problem: 1) The time delay of the sampling circuits and the PWM are neglected due to their tiny time scales [8]. 2) The dynamics of the filter in

This work is jointly supported by the National Key Research and Development Program (No. 2016YFB0900104), the National Science Foundation of China (No. 51577168) and Science & Technology Project of Yunnan Electric Power Company (yndw (2016) 000303DD00124).

Huanhai Xin, Ziheng Li, Wei Dong, Zhen Wang, and Leiqi Zhang are with the College of Electrical Engineering, Zhejiang University, Hangzhou 310027, China. Email: xinhh@zju.edu.cn.

the VFF are neglected, since the oscillation frequency under study is much lower than the cut-off frequency of the filter.

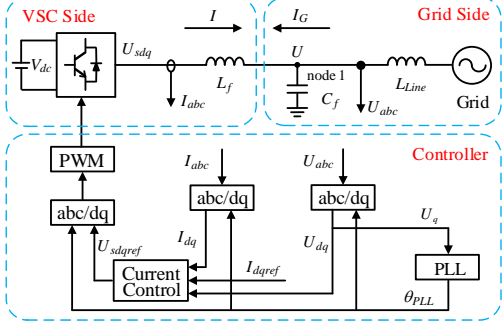

Fig. 1. Block diagram of the grid-connected converter

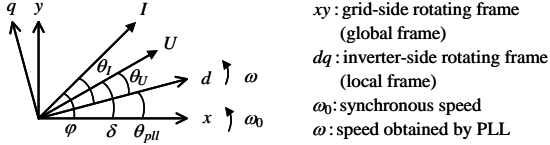

$xy$: grid-side rotating frame (global frame)
$dq$: inverter-side rotating frame (local frame)
$\omega_0$: synchronous speed
$\omega$: speed obtained by PLL

Fig. 2. Reference frames and phasor positions.

It should be emphasized that there is no constraint on the power factor (PF) of the VSC, i.e., the VSC can be the interface of a STATCOM (PF=0) or a renewable energy (usually PF ≈ 1).

### A. Dynamic Model of the VSC in Polar Coordinate

For simplicity of analysis, resistances of the filter inductor and the transmission line are neglected. In the dq frame, the dynamic model of the VSC, including the PLL, filter inductor and the current controller, can be represented by:

$$\begin{cases} \theta_{pll} = G_{pll}(s)U_q \\ U_{sdref} = (I_{dref} - I_d)G_i(s) + U_d - \omega_0 L_f I_q \\ U_{sqref} = (I_{qref} - I_q)G_i(s) + U_q + \omega_0 L_f I_d \\ U_{sd} = U_d + L_f s I_d - \omega L_f I_q \\ U_{sq} = U_q + L_f s I_q + \omega L_f I_d \end{cases} \quad (1)$$

where $G_{pll}(s)$ and $G_i(s)$ are the transfer functions of the PLL and the current controller, respectively. The meanings of other symbols are shown in Fig. 1 and Fig. 2.

To obtain the small-signal model, the following steps can be followed: 1) Linearizing (1) and solving the linearized equations, the dynamic model in the dq frame can be obtained; 2) The model is transformed to the xy frame using angle relationships shown in Fig. 2; 3) The model in the xy frame is written in the form of the polar coordinate as follows:

$$\Delta\mathbf{I} = \begin{bmatrix} \Delta I \\ I\Delta\varphi \end{bmatrix} = \begin{bmatrix} 0 & 0 \\ 0 & Y_g(s) \end{bmatrix} \begin{bmatrix} \Delta U \\ U\Delta\delta \end{bmatrix}, \quad (2)$$

where $Y_g(s) = G_i(s)G_{pll}(s)I_0 / [(G_i(s) + L_f s)(1 + G_{pll}(s)U_0)]$ and the subscript "0" denotes steady-state values. The detailed derivation is shown in Appendix A.

### B. Dynamic Model of the Grid in Polar Coordinate

According to the topology shown in Fig. 1, the model of the network can be formulated as follows:

$$\Delta\mathbf{I_G} = \begin{bmatrix} \Delta I_G \\ I_G\Delta\varphi_G \end{bmatrix} = -\Delta\mathbf{I} = -\mathbf{Y}\begin{bmatrix} \Delta U \\ U\Delta\delta \end{bmatrix} \quad (3)$$

where $\mathbf{Y} = \mathbf{Y_{Line}} + \mathbf{Y_C}$. The expressions of $\mathbf{Y_{Line}}$ and $\mathbf{Y_C}$ are shown in Appendix B. In particular, $\mathbf{Y_C}$ is 0 if the output filter only has an inductor.

## III. GENERALIZED-IMPEDANCES OF THE SYSTEM

### A. Definition of the Generalized-impedance

It follows from (2) and (3) that the VSC model and the grid model have the common structure shown as follows:

$$\begin{bmatrix} \Delta I \\ I\Delta\varphi \end{bmatrix} = \begin{bmatrix} A(s) & -C(s) \\ C(s) & D(s) \end{bmatrix} \begin{bmatrix} \Delta U \\ U\Delta\delta \end{bmatrix}. \quad (4)$$

Based on (4), the generalized-admittances ($Y_{gi}(s), i=1,2,3$) are defined as follows:

$$Y_{g1}(s) := D(s) - A(s),\ Y_{g2}(s) := A(s) + jC(s),\ Y_{g3}(s) := A(s) - jC(s). \quad (5)$$

The generalized-impedance is defined as the inverse of the generalized-admittance (i.e., $Z_{gi}(s) = Y_{gi}^{-1}(s)$).

### B. Generalized-impedances of the VSC and the Grid

It follows from (2) and (5) that the generalized-admittance of the VSC can be calculated as follows:

$$Y_{g1\_VSC}(s) = Y_g(s),\ Y_{g2\_VSC}(s) = 0,\ Y_{g3\_VSC}(s) = 0. \quad (6)$$

Note that matrix $\mathbf{Y}$ in (3) has the same structure as matrix $\begin{bmatrix} a & -b \\ b & a \end{bmatrix}$. Since $\mathbf{T}\begin{bmatrix} a & -b \\ b & a \end{bmatrix}\mathbf{T^{-1}} = \begin{bmatrix} a+jb & 0 \\ 0 & a-jb \end{bmatrix}$, it can be obtained that

$$\mathbf{TYT^{-1}} := \mathrm{diag}(Y_+, Y_-) = \mathrm{diag}(Z_+^{-1}, Z_-^{-1}), \quad (7)$$

where $\mathbf{T} = \frac{1}{\sqrt{2}}\begin{bmatrix} 1 & j \\ 1 & -j \end{bmatrix}$, $j = \sqrt{-1}$. As mentioned in Section I, matrix $\mathbf{Y}$ exists coupling terms and thus the system becomes a MIMO system.

From the definition in (5), the generalized-impedance of the grid can be obtained as follows:

$$Y_{g1\_grid}(s) = 0,\ Y_{g2\_grid}(s) = Y_+,\ Y_{g3\_grid}(s) = Y_-. \quad (8)$$

Note that $\mathbf{TYT^{-1}}$ implies the admittance is changed from the synchronous frame to the stationary frame [7], so $Y_+$ (or $Y_-$) corresponds to the positive (or negative) sequence admittance.

## IV. GENERALIZED-IMPEDANCE BASED STABILITY ANALYSIS

In this section, we derive the characteristic equations of the system, and analyze its small-signal stability.

It follows from (2) and (3) that the characteristic equation can be expressed as

$$\det\left(\begin{bmatrix} 0 & 0 \\ 0 & Y_g \end{bmatrix} + \mathbf{Y}\right) = 0, \quad (9)$$

where $\det(\cdot)$ is the determinant function.

Since $\mathbf{T}$ is invertible, (9) is equivalent to

$$\det\left(\mathbf{T}\left(\begin{bmatrix} 0 & 0 \\ 0 & Y_g \end{bmatrix} + \mathbf{Y}\right)\mathbf{T^{-1}}\right) = 0, \quad (10)$$

which is further equivalent to

$$\det\left(\frac{Y_g}{2}\begin{bmatrix} 1 & -1 \\ -1 & 1 \end{bmatrix} + \begin{bmatrix} Y_+ & 0 \\ 0 & Y_- \end{bmatrix}\right) = 0. \quad (11)$$

When $Y_g(s) = 0$, (11) implies $\det(Y_+(s)Y_-(s)) = 0$. This is a special case indicating the VSC and the network have the same resonance point, which is not the focus of this paper. Excluding this case, (11) is equivalent to

$$\det\left(\begin{bmatrix} Z_+ & -Z_- \\ -Z_+ & Z_- \end{bmatrix} + 2Z_g\mathbf{I}\right)\det\left(\frac{Y_g}{2}\begin{bmatrix} Y_+ & \\ & Y_- \end{bmatrix}\right)$$
$$=\det\left\{\begin{bmatrix} 1 & 0 \\ 1 & 1 \end{bmatrix}\left(\begin{bmatrix} Z_+ & -Z_- \\ -Z_+ & Z_- \end{bmatrix} + 2Z_g\mathbf{I}\right)\begin{bmatrix} 1 & 0 \\ -1 & 1 \end{bmatrix}\frac{Y_g\mathbf{I}}{2}\right\} = 0 \quad (12)$$

It follows from (12) that the characteristic equation of the closed-loop system can be simplified as
$$1 + Y_g(Z_+ + Z_-)/2 = 0. \quad (13)$$

From the definitions of the generalized-impedance, we can rearrange (13) as $2Z_{g1\_VSC} + Z_{g2\_grid} + Z_{g3\_grid} = 0$, which can be explained as the series circuit of the generalized-impedances of the VSC and the grid. Therefore, the oscillation of the grid-connected VSC can be regarded as the series resonance of the equivalent circuit. The system described by (13) is a SISO system, thus the complexity of analysis is reduced distinctly.

It is worth noting that the VSC and the grid are usually assumed to be stable respectively when analyzing the interaction between the VSC and the grid [6]. We also assume the VSC is stable when connected to an ideal grid, so $Y_g(s)$ and $(Z_+ + Z_-)/2$ have no pole in the right half plane. Consequently, it follows from the Nyquist criterion that

**GISC: If the Nyquist curve of $Y_g(s)(Z_+ + Z_-)/2$ does not encircle the point $(-1, j0)$, then the system is stable.**

Note that, although the form of GISC looks similar to the criterion given in [6], their meanings are different. On one hand, the open-loop transfer function used to plot the Nyquist curve is not the loop-gain $Z_{source}(s)/Z_{load}(s)$ of the source and load impedances seen at their interface, but the loop-gain of the equivalent generalized-impedance circuit. On the other hand, the criterion in [6] performs very well for single-phase system (which is a SISO system), but there may be errors resulting from the coupling terms [5] in three-phase systems. In contrast, the GISC uses the special structure of the impedance matrix in (2) and established an equivalent SISO system from a MIMO system without any approximation.

Moreover, the distance between $(-1, j0)$ and the Nyquist curve represents the stability margin, so the GISC can also be used to guide the oscillation suppression. For example, the influences of the PLL's parameters on system stability can be analyzed and tuned via the GISC. The studied system is shown in Fig. 1, in which $L_{Line}$ is set as 0.26p.u., other parameters of the VSC are listed in Table I. The Nyquist plots of the open-loop transfer function with three groups of PLL parameters, i.e., the proportional gain $K_{ppll}$ and the integral gain $K_{ipll}$, are given in Fig. 3. It shows that, in the neighborhood of the PLL parameters given in Table I, larger proportional gain ($K_{ppll}$) has more positive effect on system stability (see the curves of P2 and P3 in Fig. 3), while larger integral gain ($K_{ipll}$) has more negative effect (see the curves of P1 and P2).

## V. EXPERIMENTAL RESULTS

The hardware-in-the-loop (HIL) simulation of a three-phase grid-connected VSC is conducted in RT-LAB, where the VSC and the grid are simulated by OP5600 and the converter is controlled by a controller based on TMS320F28335. The impedance of the transmission line is changed from 0.20p.u. to 0.26p.u. at t=2s. The reference signals of the current control loop are assumed to be constant, and $I_{dref} = 1\text{p.u.}$, $I_{qref} = 0\text{p.u.}$

The stability of this system is analyzed using the GISC. The Nyquist curve of $Y_g(s)(Z_+ + Z_-)/2$ is shown in Fig. 4. The point $(-1, j0)$ is not encircled initially, thus the system is stable. With the additional inductor, the plot shows that the system has two poles in the right half plane, so the system is unstable. The HIL simulation results in Fig. 5 also verify the above analysis, where the sub-synchronous oscillation (42Hz) and super-synchronous oscillation (58Hz) occur after the impedance is increased from 0.2p.u. to 0.26p.u.

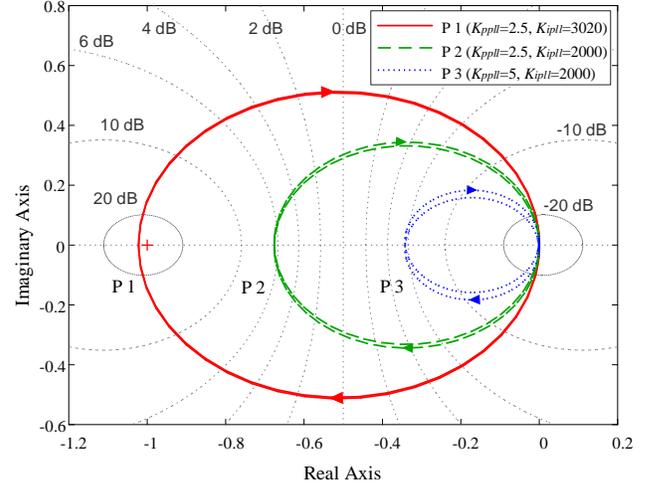

Fig. 3. Nyquist plots of the open-loop transfer function with different PLL parameters.

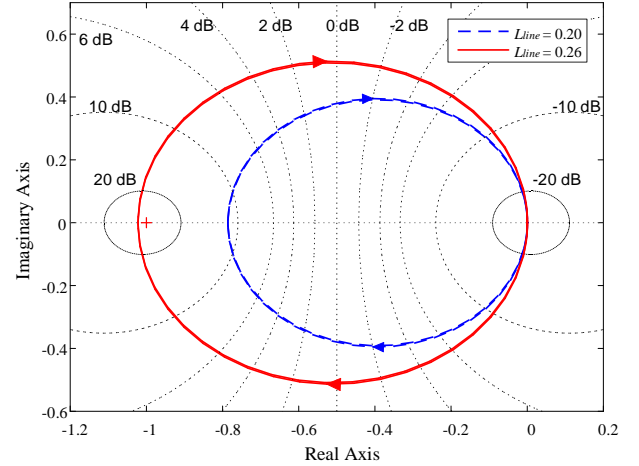

Fig. 4. Nyquist plots of the open-loop transfer function with different $L_{Line}$.

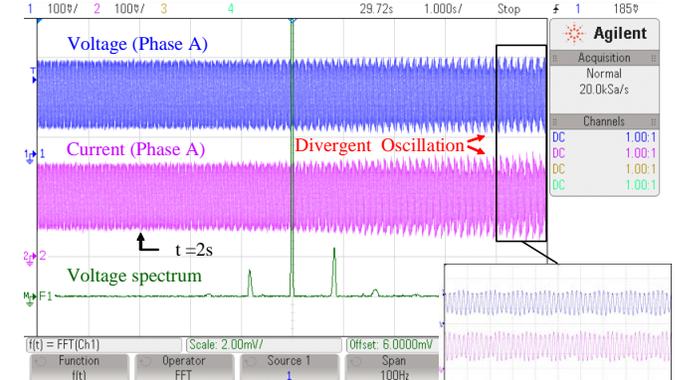

Fig. 5. Instability due to the change of inductance from 0.20p.u. to 0.26p.u.

TABLE I. PARAMETERS OF THE VSC

| Symbol | Description | Value |
|---|---|---|
| $S_b$ | Base value of power | 500kVA |
| $U_b$ | Base value of voltage | 690V |
| $L_f$ | Inductance of the inverter side filter | 0.2pu |
| $G_i(s)$ | Transfer function of the current controller | $0.6+15/s$ |
| $G_{pll}(s)$ | Transfer function of the PLL | $2.5/s+3020/s^2$ |

## VI. CONCLUSION

The generalized-impedance based stability criterion is proposed for the stability analysis of a three-phase grid-connected VSC system. By rigorous mathematical derivation, a MIMO system composed of the VSC and the grid is transformed to an equivalent SISO system composed of the generalized-impedances of the grid and the VSC. Consequently, the oscillation can be explained as the series resonance of the equivalent circuit. Such GISC can be used for the stability analysis of a three-phase system. GISC based study shows the significant effects of the PLL parameters on system stability. HIL simulation validates the effectiveness of the proposed criterion. Future work will extend this criterion to systems with multiple VSCs and use this criterion to guide VSC control design.

## APPENDIX A

By linearizing (1), the small-signal model of the inverter can be obtained in the dq frame, and can be rewritten in the form of polar coordinates as follows:

$$\begin{cases} \Delta\theta_{pll} = G_{pll}(s)U\Delta\theta_U \\ \Delta\omega = s\Delta\theta_{pll} \end{cases}, \quad (A.1)$$

$$\begin{bmatrix} \Delta U_{sdref} \\ \Delta U_{sqref} \end{bmatrix} = G_i(s)\begin{bmatrix} \Delta I_{dref} \\ \Delta I_{qref} \end{bmatrix} + (\omega_0 L_f \mathbf{T}_{\Theta 2} - G_i(s)\mathbf{T}_{\Theta 1})\begin{bmatrix} \Delta I \\ I\Delta\theta_I \end{bmatrix} + \begin{bmatrix} \Delta U \\ U\Delta\theta_U \end{bmatrix}, \quad (A.2)$$

$$\begin{bmatrix} \Delta U_{sd} \\ \Delta U_{sq} \end{bmatrix} = \begin{bmatrix} \Delta U \\ U\Delta\theta_U \end{bmatrix} + L_f s G_{pll}(s)\begin{bmatrix} 0 & -I_q \\ 0 & I_d \end{bmatrix}\begin{bmatrix} \Delta U \\ U\Delta\theta_U \end{bmatrix} + \omega_0 L_f \mathbf{T}_{\Theta 2}\begin{bmatrix} \Delta I \\ I\Delta\theta_I \end{bmatrix} + L_f s\mathbf{T}_{\Theta 1}\begin{bmatrix} \Delta I \\ I\Delta\theta_I \end{bmatrix}, \quad (A.3)$$

where $\mathbf{T}_{\Theta 1} = \begin{bmatrix} \cos\theta_I & -\sin\theta_I \\ \sin\theta_I & \cos\theta_I \end{bmatrix}$ and $\mathbf{T}_{\Theta 2} = \begin{bmatrix} -\sin\theta_I & -\cos\theta_I \\ \cos\theta_I & -\sin\theta_I \end{bmatrix}$.

It follows from the assumptions that $\Delta U_{sdref} = \Delta U_{sd}$, $\Delta U_{sqref} = \Delta U_{sq}$, and $\Delta I_{dref} = \Delta I_{qref} = 0$ are satisfied. Thus, from (A.2) and (A.3) the following equation is obtained:

$$0 = sG_{pll}(s)L_f U\Delta\theta_U\begin{bmatrix} -I_q \\ I_d \end{bmatrix} + (sL_f + G_i(s))\mathbf{T}_{\Theta 1}\begin{bmatrix} \Delta I \\ I\Delta\theta_I \end{bmatrix}. \quad (A.4)$$

Considering that the angles in (A.4) satisfy $\Delta\theta_U = \Delta\delta - \Delta\theta_{pll}$ and $\Delta\theta_I = \Delta\varphi - \Delta\theta_{pll}$ (shown in Fig. 2), $\Delta\theta_{pll}$ in (A.1) can be further expressed as $\Delta\theta_{pll} = G_{pll}(s)U\Delta\delta/(1+G_{pll}(s)U)$ in terms of $\Delta\delta$. Thus, by eliminating $\Delta\theta_U$, $\Delta\theta$, and $\Delta\theta_{pll}$ in (A.4), (A.5) holds:

$$\begin{bmatrix} \Delta I \\ I\Delta\varphi \end{bmatrix} = \begin{bmatrix} 0 & 0 \\ 0 & Y_g(s) \end{bmatrix}\begin{bmatrix} \Delta U \\ U\Delta\delta \end{bmatrix}, \quad (A.5)$$

where $Y_g(s) = G_i(s)G_{pll}(s)I_0 / [(G_i(s)+L_f s)(1+G_{pll}(s)U_0)]$ and the subscript "0" denotes steady-state values.

## APPENDIX B

The linearized state equation of an inductor in xy reference frame can be expressed as:

$$\begin{bmatrix} \Delta I_{kx} \\ \Delta I_{ky} \end{bmatrix} = \frac{1}{L_k s^2 + L_k \omega^2}\begin{bmatrix} s & \omega \\ -\omega & s \end{bmatrix}\left\{\begin{bmatrix} \Delta U_{ix} \\ \Delta U_{iy} \end{bmatrix} - \begin{bmatrix} \Delta U_{jx} \\ \Delta U_{jy} \end{bmatrix}\right\}, \quad (B.1)$$

where $i$ and $j$ are the nodes next to the inductor.

Since the voltage of the ideal grid is considered to be constant, the linearized state equation of $L_{Line}$ can be expressed as:

$$\begin{bmatrix} \Delta I_{Linex} \\ \Delta I_{Liney} \end{bmatrix} = \frac{1}{L_{Line} s^2 + L_{Line}\omega^2}\begin{bmatrix} s & \omega \\ -\omega & s \end{bmatrix}\begin{bmatrix} \Delta U_x \\ \Delta U_y \end{bmatrix}. \quad (B.2)$$

Denote $\phi_{Line}$ to be the power factor angle at the node 1 of $L_{Line}$. Then (B.2) can be converted to the polar coordinates as follows:

$$\begin{bmatrix} \Delta I_{Line} \\ I_{Line}\Delta\varphi_{Line} \end{bmatrix} = \mathbf{Y}_{Line}\begin{bmatrix} \Delta U \\ U\Delta\delta \end{bmatrix}, \quad (B.3)$$

where $\mathbf{Y}_{Line} = \frac{1}{L_{Line} s^2 + L_{Line}\omega^2}\begin{bmatrix} s & \omega \\ -\omega & s \end{bmatrix}\begin{bmatrix} \cos\phi_{Line} & -\sin\phi_{Line} \\ \sin\phi_{Line} & \cos\phi_{Line} \end{bmatrix}$.

Similarly, the linearized state equation of a capacitor in xy coordinates can be expressed as:

$$\begin{bmatrix} \Delta I_{kx} \\ \Delta I_{ky} \end{bmatrix} = C_k \begin{bmatrix} s & -\omega \\ \omega & s \end{bmatrix}\left\{\begin{bmatrix} \Delta U_{ix} \\ \Delta U_{iy} \end{bmatrix} - \begin{bmatrix} \Delta U_{jx} \\ \Delta U_{jy} \end{bmatrix}\right\}, \quad (B.4)$$

where $i$ and $j$ are the nodes next to the inductor.

Denote $\phi_C$ to be the power factor angle at the node 1 of $C_f$. Since the voltage of the ground is also constant, the linearized state equation of $C_f$ can be expressed as:

$$\begin{bmatrix} \Delta I_C \\ I_C\Delta\varphi_C \end{bmatrix} = \mathbf{Y}_C \begin{bmatrix} \Delta U \\ U\Delta\delta \end{bmatrix}, \quad (B.5)$$

where $\mathbf{Y}_C = C_C \begin{bmatrix} s & -\omega \\ \omega & s \end{bmatrix}\begin{bmatrix} \cos\phi_C & -\sin\phi_C \\ \sin\phi_C & \cos\phi_C \end{bmatrix}$.

## REFERENCES

[1] F. Blaabjerg, Z. Chen, and S. B. Kjaer, "Power electronics as efficient interface in dispersed power generation systems," *IEEE Trans. Power Electron.*, vol. 19, no. 5, pp. 1184-1194, Sep. 2004.

[2] B. Wen, D. Boroyevich, R. Burgos, P. Mattavelli, and Z. Shen, "Analysis of D-Q small-signal impedance of grid-tied inverters," *IEEE Trans. Power Electron.*, vol. 31, no. 1, pp. 675-687, Jan. 2016.

[3] B. Wen, D. Dong, D. Boroyevich, R. Burgos, P. Mattavelli, and Z. Shen, "Impedance-based analysis of grid-synchronization stability for three-phase paralleled converters," *IEEE Trans. Power Electron.*, vol. 31, no. 1, pp. 26-37, Jan. 2016.

[4] M. Cespedes, J. Sun, "Impedance modeling and analysis of grid-connected voltage-source converters," *IEEE Trans. Power Electron.*, vol. 29, no. 3, pp. 1254-1261, Mar. 2014.

[5] M. K. Bakhshizadeh, X. Wang, F. Blaabjerg, et. al., "Couplings in phase domain impedance modeling of grid-connected converters," *IEEE Trans. Power Electron.*, vol. 31, no. 10, pp. 6792-6796, Oct. 2016.

[6] J. Sun, "Impedance-based stability criterion for grid-connected inverters," *IEEE Trans. Power Electron.*, vol.26, no.11, pp.3075-3078, Nov. 2011.

[7] A. Rygg, M. Molinas, Z. Chen, and X. Cai, "A modified sequence domain impedance definition and its equivalence to the dq-domain impedance definition for the stability analysis of ac power electronic systems," *IEEE J. Emerging Sel. Topics Power Electron.*, vol. 4, no. 4, pp. 1383-1396, Dec. 2016.

[8] M. Zhao, X. Yuan, J. Hu, and Y. Yan, "Voltage dynamics of current control time-scale in a VSC-connected weak grid," *IEEE Trans. Power Syst.*, vol. 31, no.4, pp. 2925-2937, Jul. 2016.